\documentclass[final,5p,times,twocolumn]{elsarticle}

\usepackage{graphicx}
\usepackage{amssymb}
\usepackage{subfigure}

\journal{Nuclear Instruments and Methods A}

\begin{document}

\begin{frontmatter}

\title{Performance of novel silicon n-in-p planar Pixel Sensors}

\author[A]{C. Gallrapp}
\author[A]{A. La Rosa\corref{cor1}}
\ead{alessandro.larosa@cern.ch}
\author[B]{A. Macchiolo}
\author[B]{R. Nisius}
\author[A]{H. Pernegger}
\author[C]{R.H.  Richter}
\author[B]{P. Weigell}

\address[A]{CERN, Geneva 23, CH-1211, Switzerland}
\address[B]{Max-Planck-Institut f\"ur Physik (Werner-Heisenberg-Institut) F\"ohringer Ring 6, D-80805 M\"unchen, Germany}
\address[C]{Max-Planck-Institut Halbleiterlabor, Otto Hahn Ring 6, D-81739 M\"unchen, Germany}

\cortext[cor1]{Corresponding author, now at Section de Physique (DPNC), Universit\'e\\ de Gen\`eve, 24 quai Ernest Ansermet 1211 Gen\`eve 4, Switzerland.}


\begin{abstract}
The performance of novel n-in-p planar pixel detectors, designed for future upgrades of the ATLAS Pixel system is presented. The n-in-p silicon sensors technology is a promising candidate for the pixel upgrade thanks to its radiation hardness and cost effectiveness, that allow for enlarging the area instrumented with pixel detectors. The n-in-p modules presented here are composed of pixel sensors produced by CiS connected by bump-bonding to the ATLAS readout chip FE-I3.\\ 
The characterization of these devices has been performed  before and after irradiation up to a fluence of 5\,x\,$10^{15}$\,1\,MeV\,n$_{\mathrm{eq}}$cm$^{-2}$. Charge collection measurements carried out with radioactive sources have proven the functioning of this technology up to these particle fluences. First results from beam test data with a 120 GeV/c pion beam at the CERN-SPS are also discussed, demonstrating a high tracking efficiency of  (98.6\,$\pm$\,0.3)\,\%  and a high collected charge of about 10\,ke for a device irradiated at the maximum fluence and biased at 1\,kV.
\end{abstract}

\end{frontmatter}

\section{Introduction}
\label{sec:Intro}

The upgrade of the Large Hadron Collider (LHC), known as the High Luminosity-LHC or HL-LHC,  is planned to be achieved in two steps \cite{LHC}. From 2016 the luminosity is foreseen to be increased to around (2-3)\,x\,$10^{34}$\,cm$^{-2}$s$^{-1}$. Focusing on the ATLAS tracker, an upgrade of the pixel system, called NewPix, is under discussion for the year 2018 \cite{NewPix} when, depending on the operational performance, a replacement and extension of the entire pixel detector may be needed.  After 2020, a major upgrade will increase the luminosity further to 5\,x\,$10^{34}$\,cm$^{-2}$s$^{-1}$, and requires the replacement of the ATLAS strip tracker and of the pixel system, in case the latter will not have been already exchanged in 2018.  In this scenario, the innermost layers of the ATLAS vertex detector system will have to sustain very high fluences of more than $10^{16}$ 1\,MeV equivalent neutrons per square centimeter (n$_{\mathrm{eq}}$cm$^{-2}$) with 3000\,fb$^{-1}$ total integrated luminosity at the end of the LHC lifetime around 2030.
The total surface covered by pixel sensors is foreseen to increase from the present 1.8\,m$^{2}$ to about 10\,m$^{2}$ in the upgraded ATLAS detector. This significantly larger area requires the use of cost-effective n-in-p sensors instead of the standard n-in-n technology used to instrument the present pixel detectors of the LHC experiments.  
To investigate the performance and the range of fluences for which n-in-p pixel sensors can be used, a common production within the CERN-RD50 Collaboration \cite{RD50} and the ATLAS Upgrade Planar Pixel Sensor R\&D group has been realized. Preliminary results from this production have already been presented in \cite{{RESMDD10},{TIPP2011}}.

\section{Sensor description}
\label{sec:sensor}

\begin{figure}[h!]
\centering
\includegraphics[scale=0.5]{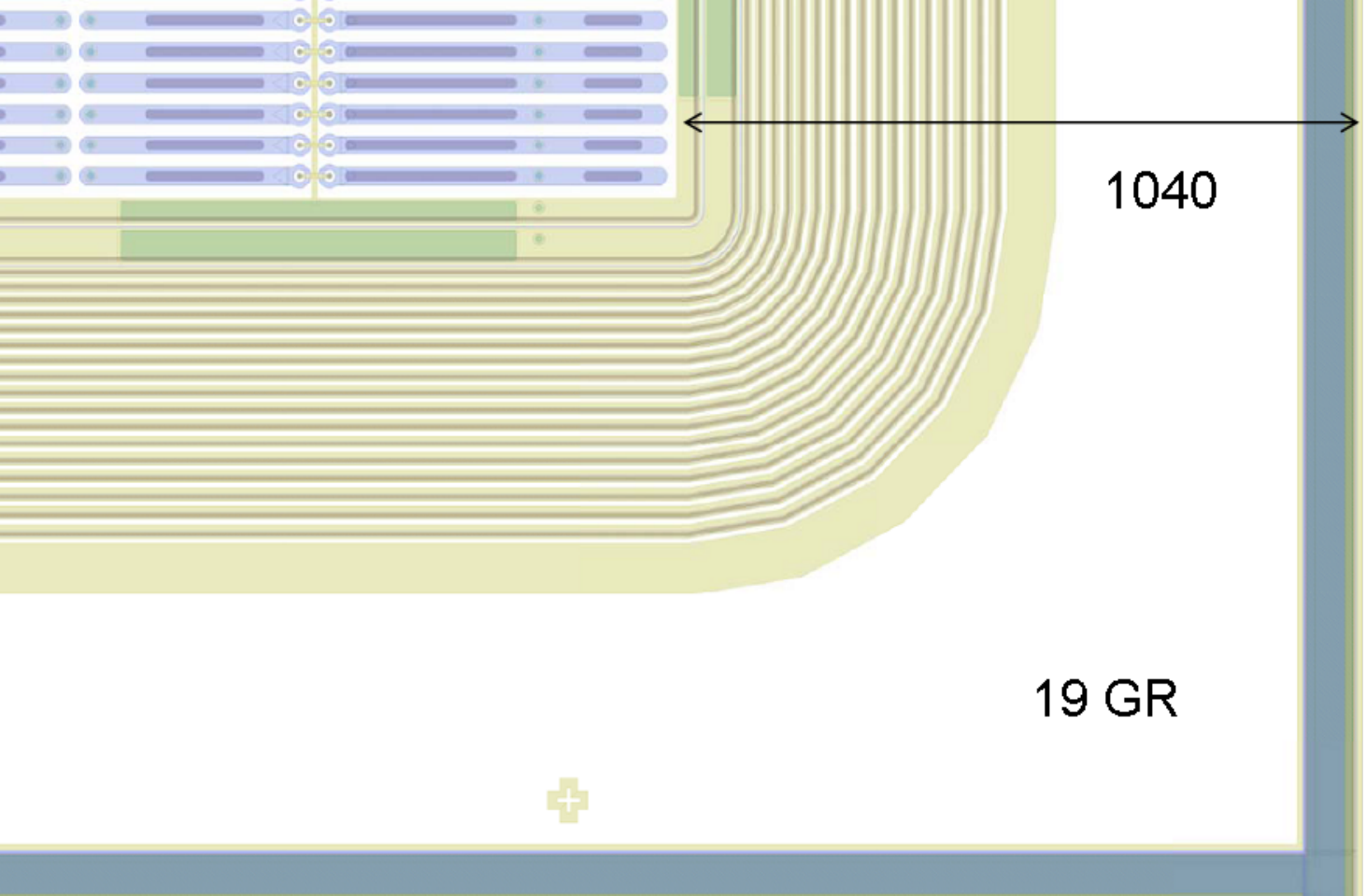}
\caption{Design of the 19 guard-ring (GR) structure on the front-side of the n-in-p pixel sensor.}
\label{fig:1040}
\end{figure}

\begin{figure}[tbp!]
\centering
\includegraphics[scale=0.35]{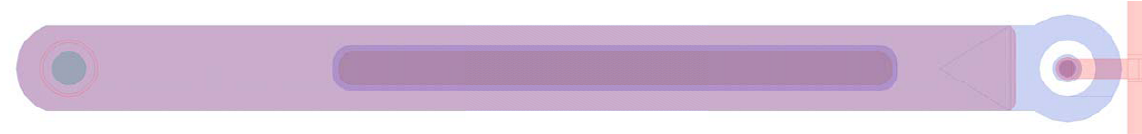}
\caption{
Design of a single pixel. The implantation extends over the entire structure shown, and has a ring shaped opening at the punch through bias dot displayed on the right side.
The metal layer, covering most of the implant, is shown as a large rectangle with rounded corners on the left side. The T-shaped structure at the far right end comprises the metal lines, connecting the bias dot to the bias ring.
The opening in the nitride and oxide layers is displayed as the rectangle in the centre of the pixel.
The small circle at the left end of the pixel is the opening in the passivation, where the pixel will be connected with bump bonding.
}
\label{fig:pixel-layer}
\end{figure}

The production of n-in-p pixel sensors has been performed by CiS (Erfurt, Germany) \cite{CiS} on 4'' wafers of high resistivity ($\rho$ $>$10 k$\Omega$cm) Float Zone material, with a thickness of 300\,$\mu$m. The n-in-p technology with its guard-ring structure implemented on the front-side, shown in Figure\,\ref{fig:1040}, allows for a single sided processing of the wafer. This implies a reduced number of process steps, leading to a cost reduction. The design of the pixel cells and the guard-ring structure was first implemented in a n-in-p pixel production \cite{Pixel2010} by MPP-HLL. To achieve a narrower inactive region, two modified versions of the guard-ring structure with 15 and 19 rings were implemented in the CiS production.
\\
The biasing of the pixel matrix is realized via the punch-through structure of each individual pixel cell, connected with aluminum lines to the common bias ring. This allows for the electrical characterization of the sensor before the interconnection to the front-end electronic chip. Since, as shown in Figure\,\ref{fig:pixel-layer}, the central dot of the n$^{+}$ implantation is not directly connected to the pixel structure, a reduced charge collection efficiency in this region is expected.\\
Each wafer hosts ten FE-I3 \cite{FEI3} compatible pixel sensors, which consist of normal, long, ganged and inter-ganged pixels arranged in 18 columns and 160 rows, having a total active surface of  7.7\,mm\,$\times$\,8.2\,mm. The normal pixels cover the central region between the second and the 17$^{th}$ column and extended from row zero to 152, with a pixel size of 400\,$\mu$m\,x\,50\,$\mu$m. The first and the last column consist of long pixels measuring 600\,$\mu$m\,x\,50\,$\mu$m.
The uppermost seven rows are populated alternating with ganged and inter-ganged pixels, in which the inter-ganged pixels have the same layout as the normal pixels. The ganged pixels assure the connection to four additional pixel rows, which are not bump-bonded to the front-end chip. This implies that the ganged pixel have double the size of a normal pixel cell.\\
Before the deposition of the Under Bump Metallization (UBM), the sensors front side has been passivated with a BCB layer (BenzoCycloButen, Cycloten) that provides higher electrical insulation capability than SiO$_{2}$. This ensures the severe insulation requirements of n-in-p pixels, where the area on the front-side, outside the guard-ring structures, is kept at a large potential, while facing the read-out chip, kept at ground potential.\\
The thickness of the deposited BCB on the sensor surface is 3\,$\mu$m and this passivation layer has only been opened in correspondence of the UBM pads. Since the process has been performed at wafer level the dicing lines and the vertical sides of the sensors are not protected in this approach.  The chip lies inside the sensor guard-ring structures everywhere except for the side where the chip balcony with the EOC (End Of Columns) logic is located.\\

\section{Modules and experimental setups}
\label{sec:samples-expsetup}

\begin{figure}[t]
	\centering
	\includegraphics[scale=0.06]{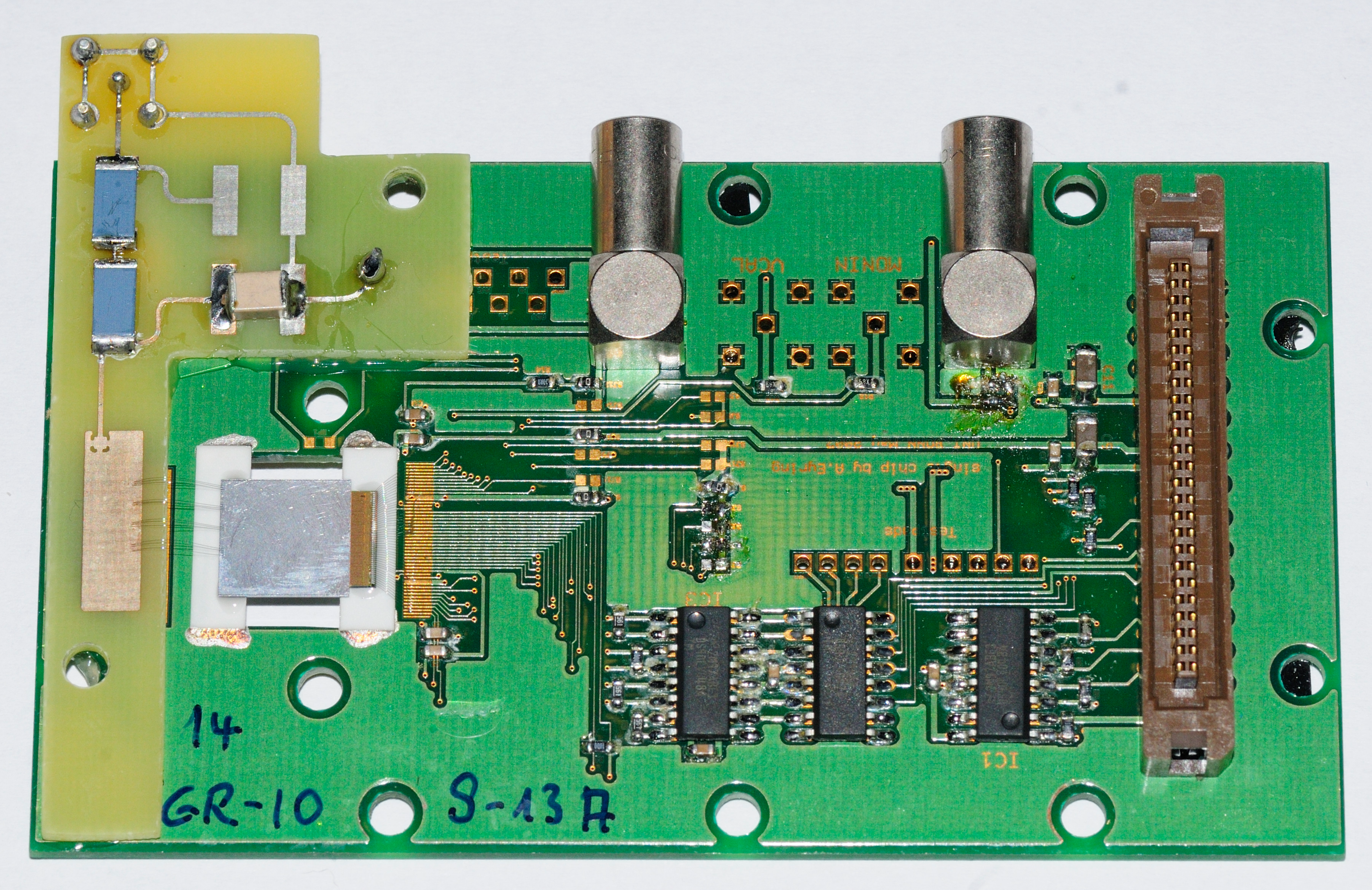}
	\caption{Photograph of a single chip module on a test card. The module can be seen on the left hand side.}
	\label{fig:SCA}
\end{figure}

A single chip module (or module) is built by the FE-I3 front-end chip \cite{FEI3}  bump-bonded to a sensor, and it is used for the characterization and qualification of the sensor technology. Figure~\ref{fig:SCA} shows a single chip module on a test card which provides the necessary connectors for communication and power supply.\\
Wire bonds are used to connect the front-end chip with the test card, and to apply the bias voltage for depleting the sensors. The module is glued onto two ceramic stripes, providing mechanical support, over an opening in the test card, which allows for source measurements with an external trigger system. Each pixel is bump-bonded to a corresponding front-end pixel cell, which consists of a digital and analog part. Signals from the sensor are first processed in the analog part, where two amplifiers determine whether the signal is higher than the threshold, and perform the shaping of the feedback current. The output is passed on to the digital part, where a discriminator converts the analog signal, decoded in counts of an external clock running at 40\,MHz. This digital signal is the time over threshold (ToT), which can be converted into the collected charge from the sensor. The front-end chip also allows to use the hit pixel as an internal trigger. 
This is especially useful for $\gamma$-source measurements, that are used to compare the measured charge with the expected deposited charge, calculated from the known photon energies.\\
To perform accelerated aging and long term viability studies, a collection of several modules has been irradiated at a broad range of fluences, with different particles, up to a fluence of 5x$10^{15}\,$n$_{\mathrm{eq}}$cm$^{-2}$. Several modules have been irradiated with neutrons at the TRIGA reactor at the Josef Stefan Institut (JSI) \cite{Snoj}, and others at the Karlsruhe Institute of Technology (KIT) \cite{KIT} with a 25\,MeV proton beam.
The modules have been stored in a cold environment after the irradiation, to prevent annealing. However, the modules irradiated at JSI had to be dismounted from the test card for irradiation and remounted afterwards, and thus an annealing of 1-2 days could not be avoided.
Table \ref{tab:irrad_devices} lists the irradiated assemblies with their respective target fluences, scaled to 1\,MeV\,n$_{\mathrm{eq}}$, following  the NIEL hypothesis \cite{ROSE}.
\begin{table}[h!]
\begin{center}
\begin{tabular}{|p{2cm}|*{3}{c|}}
\hline
Module ID & Fluence [n$_{\mathrm{eq}}$cm$^{-2}$] & Facility \\
\hline
A	&	1x$10^{15}$	&	KIT	\\
B	&	1x$10^{15}$	&	JSI	\\
C	&	2x$10^{15}$	&	JSI	\\
D	&	3x$10^{15}$	&	JSI	\\
E	&	5x$10^{15}$	&	JSI	\\
\hline
\end{tabular}
\end{center}
\caption{Overview of irradiated modules with the respective target fluences. }
\label{tab:irrad_devices}
\end{table}
\\
The modules have been characterized by measuring the leakage current versus the reverse bias voltage, the threshold, the noise, and the response to radioactive $\gamma$- and $\beta$-sources. 
These measurements have been performed in a climate chamber at low humidity and stable air temperature.
A selection of representative results,  obtained before and after irradiation, are reported in Section\,4.\\
The performance of a number of irradiated modules have been investigated at the CERN SPS North Area with a 120 GeV/c $\pi$$^{+}$ beam. Thanks to the high momentum of the beam particles, the multiple scattering is minimized, leading to high precision tracking measurements, performed with the EUDET beam telescope \cite{EUDET_2}. 
To avoid annealing during the data taking, and to ensure low leakage current, the modules under test have been cooled to negative Celsius temperatures using a dry-ice system. 
The results from the beam test analysis are reported in Section\,5, concentrating on the tracking efficiency, the charge sharing, the cluster size, and the charge collection across the pixel area.\\

\section{Characterization and source-test measurements}
\label{sec:lab}
\subsection{IV measurements}
\begin{figure}[htb]
\centering
\includegraphics[scale=0.4]{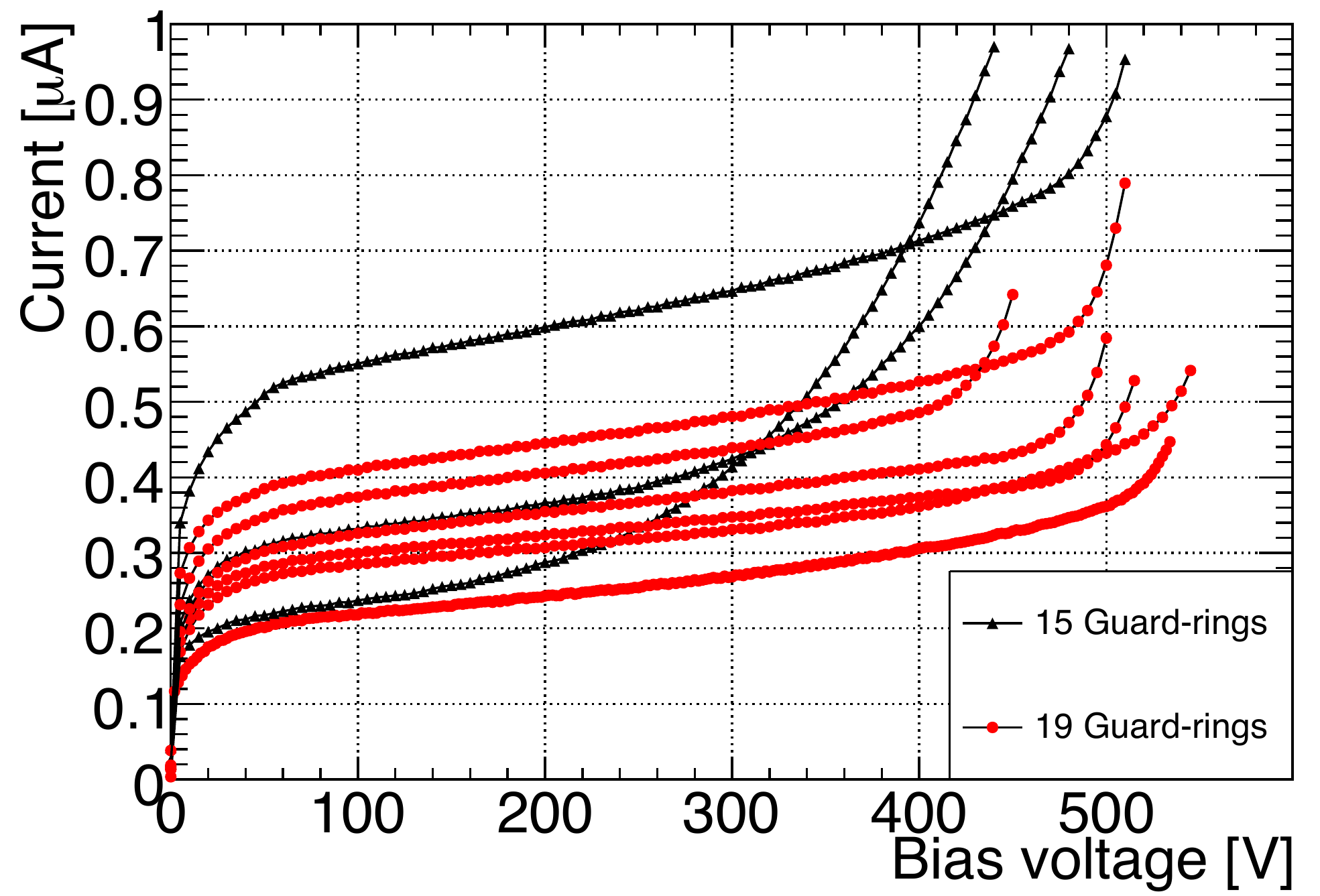}
\caption{IV characteristics of selected non-irradiated sensors after under bump metallization performed at a temperature of $20\,^{\circ}\mathrm{C}$ controlled in the probe station.}
\label{fig:IVunirrad}
\end{figure}

\begin{figure}[htb]
\centering
\includegraphics[scale=0.4]{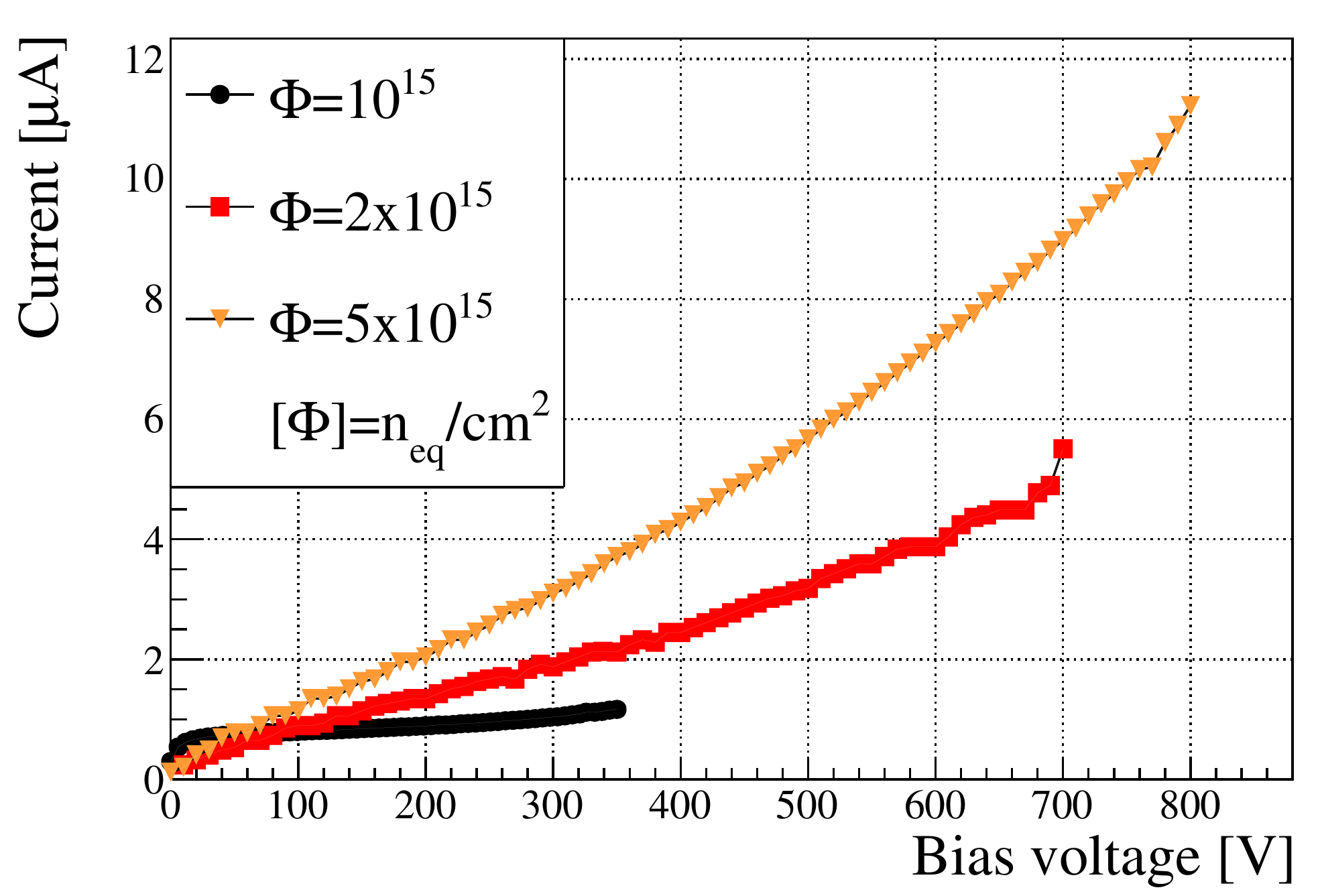}
\caption{IV characteristics of the selected modules:  B, C and E, irradiated up to 1, 2 and 5\,x\,10$^{15}$\,n$_{\mathrm{eq}}$cm$^{-2}$ respectively. The measurements have been scaled to a temperature of $-20\,^{\circ}\mathrm{C}$ \cite{IVscaling}. }
\label{fig:IVirrad}
\end{figure}

The leakage current measurement will reveal any damage coming from dicing and the flip-chipping procedure. It is also used to evaluate the irradiation fluence, and the annealing condition of irradiated assemblies.\\
All the manufactured sensors were electrically characterized before being selected for flip-chipping to the front-end chip. Figure \ref{fig:IVunirrad} shows the IV characteristics at room temperature, of the measured sensors after UBM treatment. All selected samples show leakage currents below 0.6\,$\mu$A,  when operated at a bias voltage of 150\,V. In most sensors, the breakdown takes place at voltages above 400\,V, while the sensors are already fully depleted at about 60\,V.\\
Figure \ref{fig:IVirrad} shows the IV characteristics of the samples irradiated with neutrons up to 1, 2 and 5\,x\,10$^{15}$\,n$_{\mathrm{eq}}$cm$^{-2}$. As expected, the breakdown voltage of the irradiated sensors shifts to higher values, and for the fluence of 5\,x\,10$^{15}$\,n$_{\mathrm{eq}}$cm$^{-2}$ exceeds 800\,V. 
Also the leakage currents are in agreement with expectations, showing increasing leakage currents with increasing fluences.
\subsection{Response to radioactive sources}
\begin{figure}[h!]
\centering
\subfigure[]{
\includegraphics[scale=0.45]{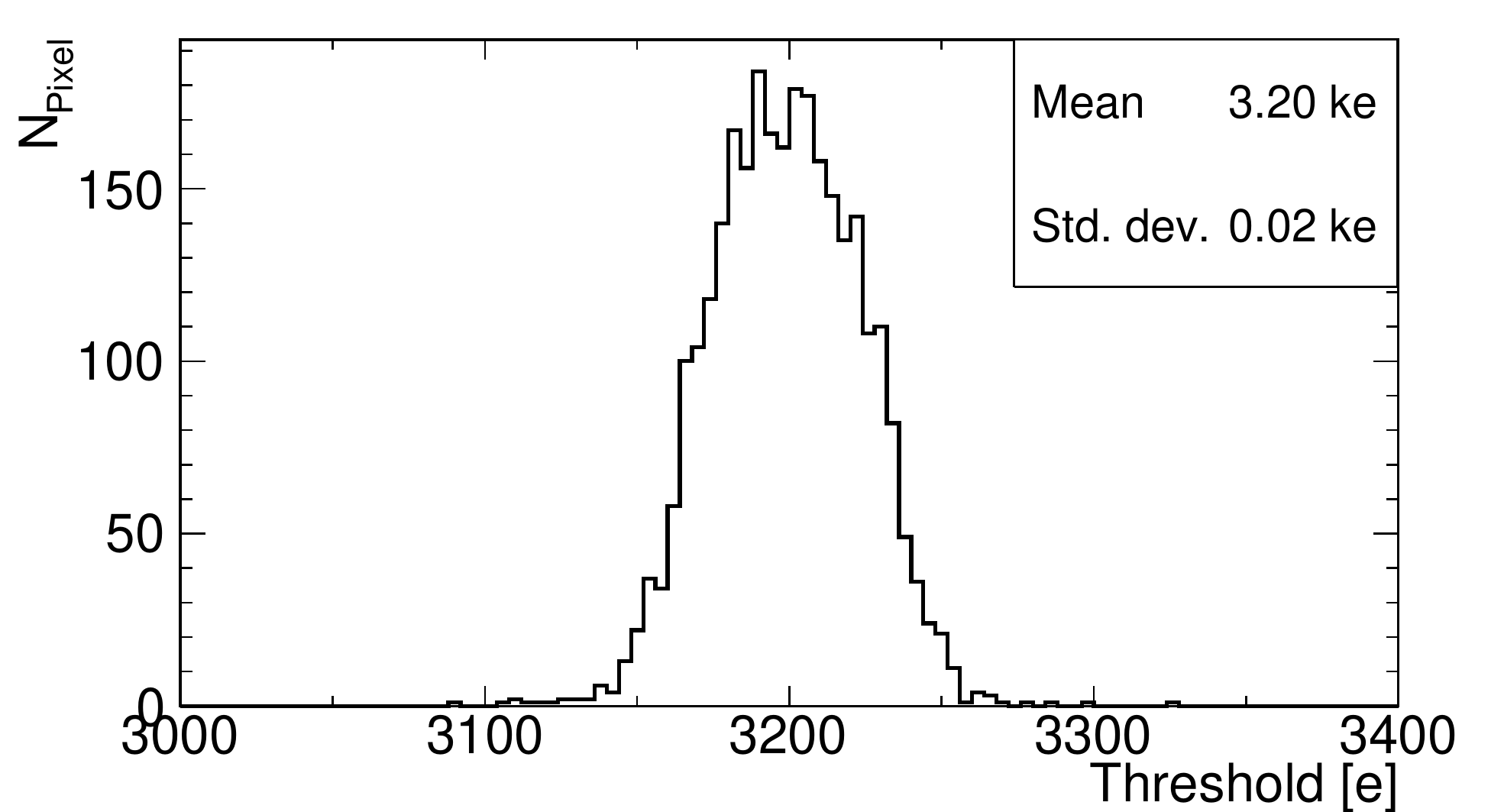}
\label{fig:unirrad-th}
}
\subfigure[]{
\includegraphics[scale=0.42]{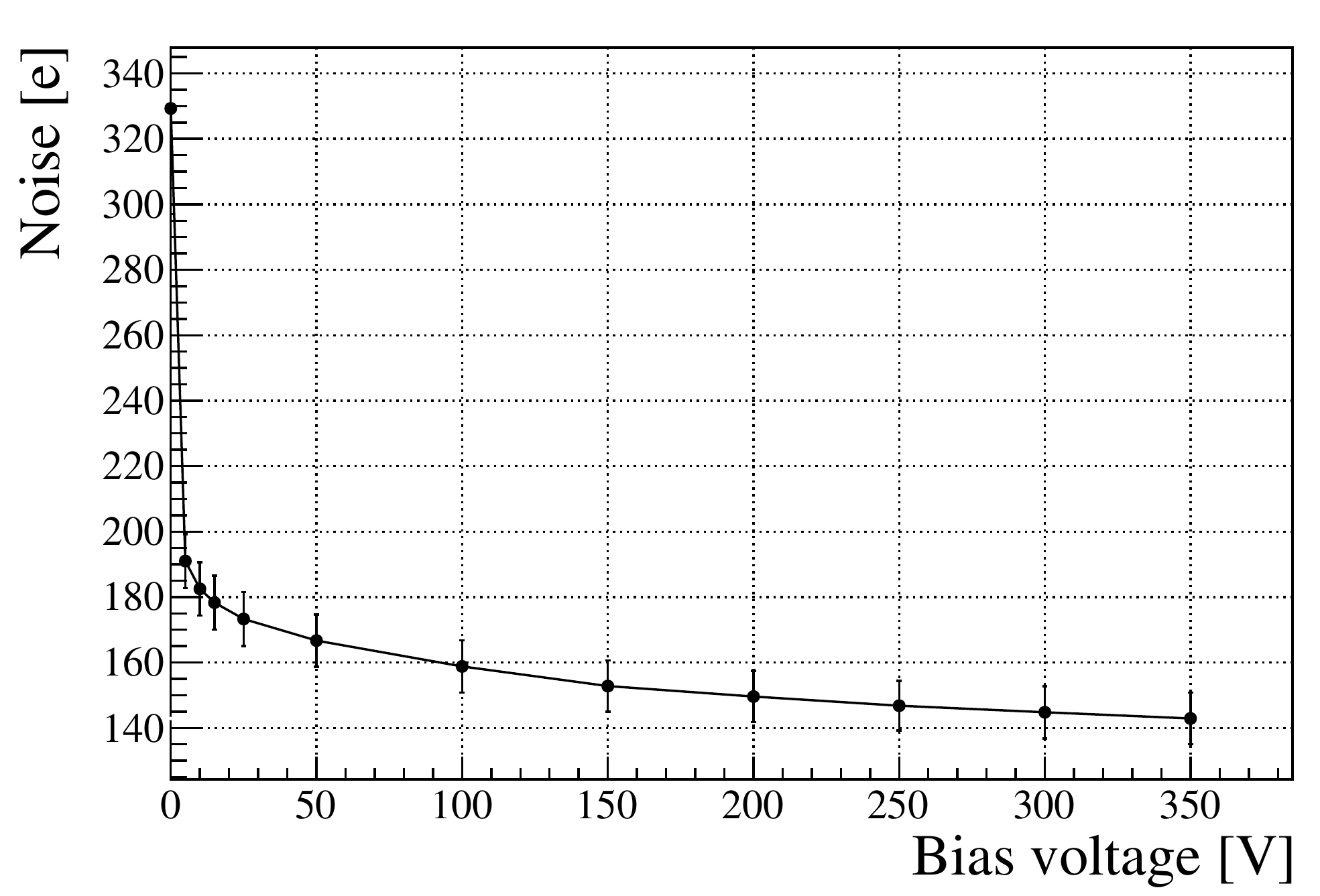}
\label{fig:unirrad-noise_hv}
}
\caption{\subref{fig:unirrad-th} Threshold distribution of a non-irradiated sample after tuning to a target threshold value of 3200\,e. \subref{fig:unirrad-noise_hv} Noise measurements as a function of the applied bias voltage measured on a non-irradiated sample.}
\end{figure}

Measurements with radioactive sources are a possibility to analyze the sensor performance of devices before and after irradiation. A common way to perform measurements is to use a $\gamma$-source as $^{241}$Am 
to verify the integrity of the bump-bond connections between the sensor and the front-end chip, and a $^{90}$Sr $\beta$-source, to study the charge collection in the sensor.\\
Prior to the source tests, the front-end chip has been tuned to a threshold of 3.2\,ke. A typical threshold distribution for a non-irradiated sample is shown in Figure \ref{fig:unirrad-th}. 
Figure \ref{fig:unirrad-noise_hv} shows the noise as a function of the bias voltage, which agrees with the fact that the main contribution to sensor noise for non-irradiated sensors is known to come from the sensor capacitance. At a bias voltage of 150\,V, the noise lies in the range from 145\,e to 160\,e, which is compatible with what has been measured for the n-in-n modules \cite{atlaspixel}. Very uniform results have been obtained for all modules that have been tested.
\begin{figure}[h]
\centering
\subfigure[]{
\includegraphics[scale=0.45]{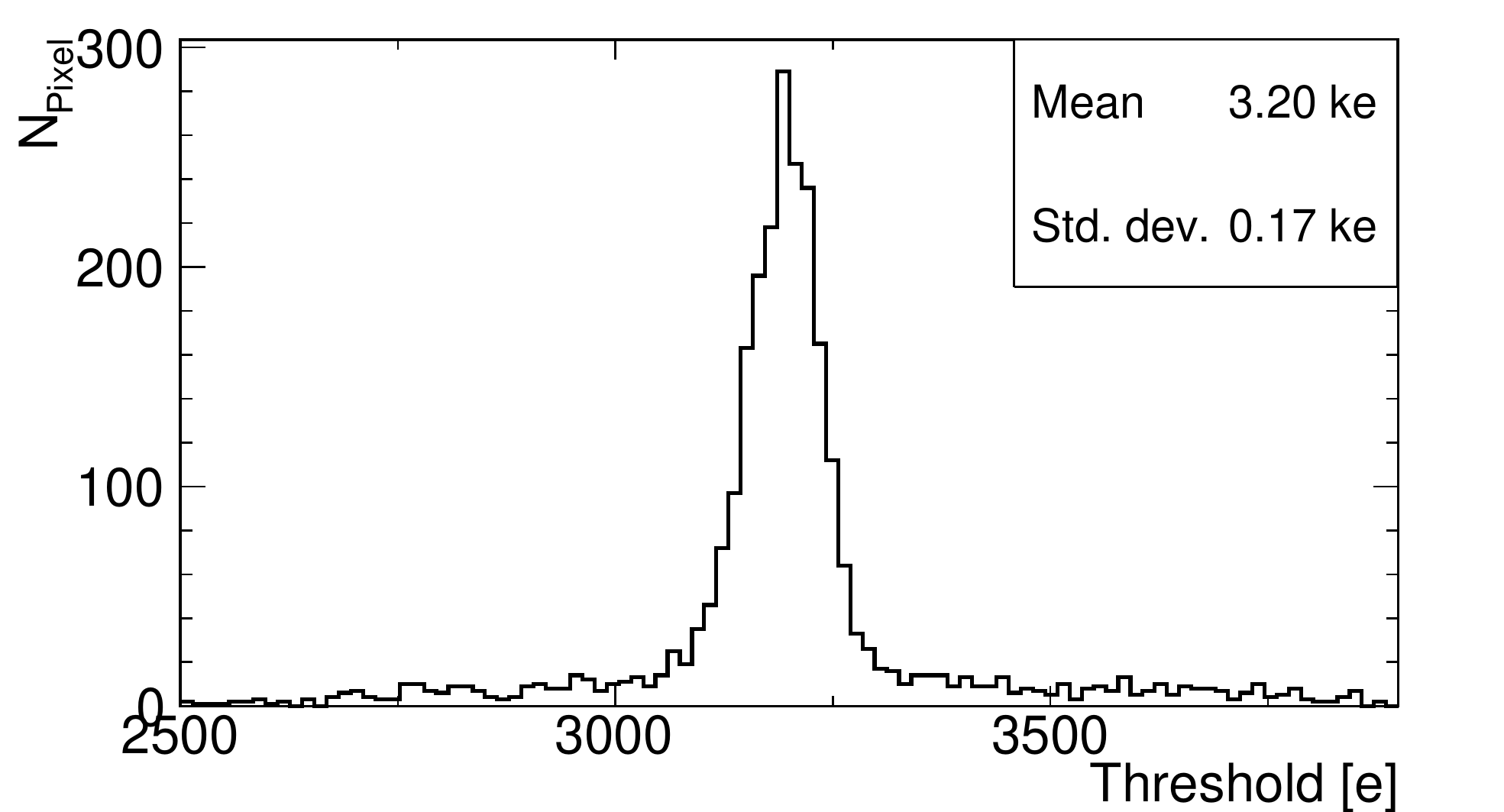}
\label{fig:irrad-th}
}
\subfigure[]{
\includegraphics[scale=0.45]{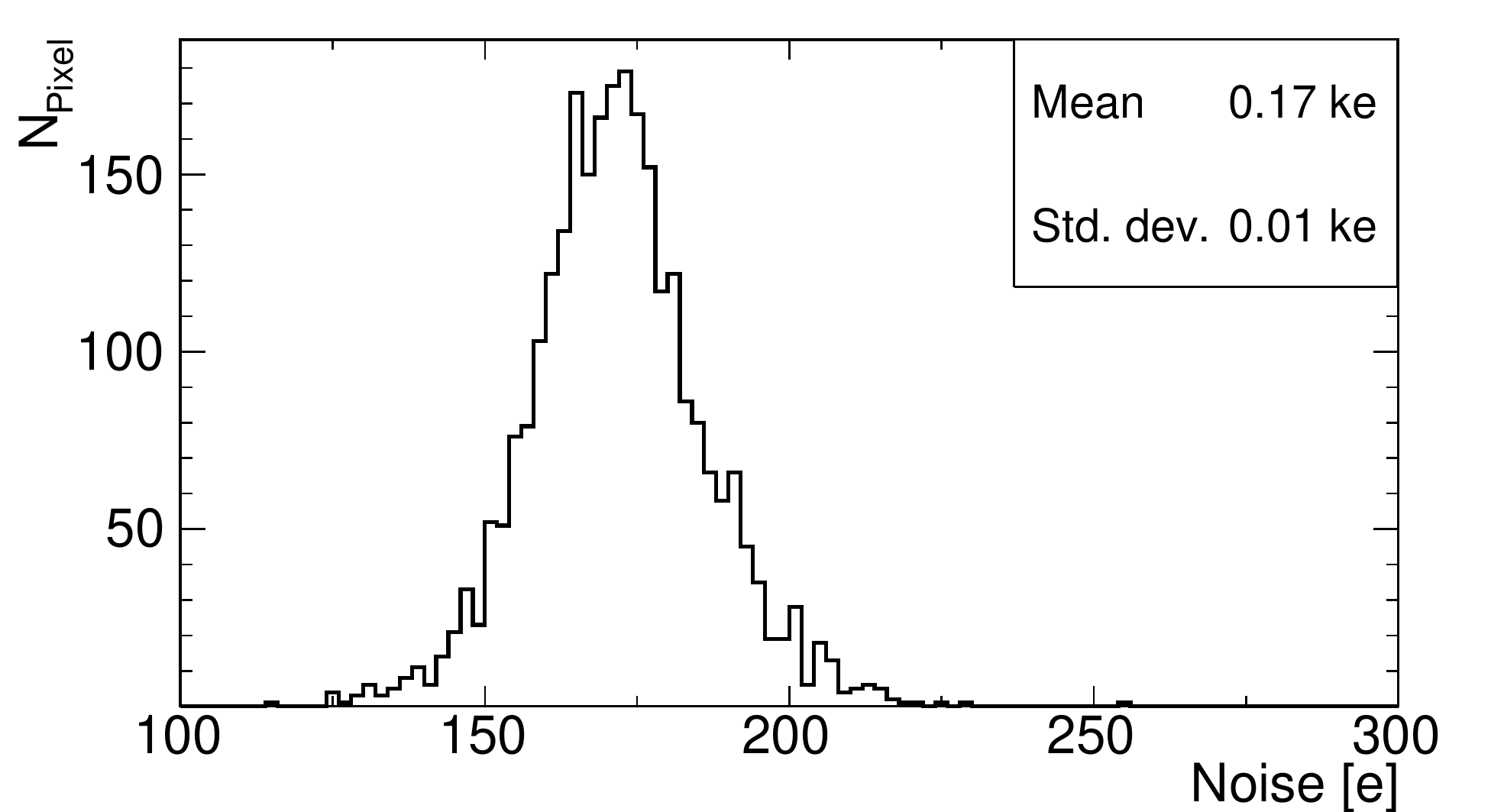}
\label{fig:irrad-noise}
}
\caption{\subref{fig:irrad-th} Threshold and \subref{fig:irrad-noise} noise distributions of the sample E, irradiated to 5\,x\,10$^{15}$\,n$_{\mathrm{eq}}$cm$^{-2}$, biased at 1000\,V, and kept at an air temperature of $-60\,^{\circ}\mathrm{C}$. }
\end{figure}
\\
In figure \ref{fig:irrad-th} and \ref{fig:irrad-noise} the threshold and noise distributions for sample E, irradiated up to 5\,x\,10$^{15}$\,n$_{\mathrm{eq}}$cm$^{-2}$, are shown at a bias voltage of 1000V and for an air temperatures in the climate chamber of $-60\,^{\circ}\mathrm{C}$. In these extreme conditions, far beyond the FE-I3 specifications \cite{FEI3}, the module is more difficult to tune. Nonetheless, three quarters of the pixels could be tuned to the desired threshold of 3.20\,ke with a standard deviation of 0.17\,ke. Figure \ref{fig:irrad-noise} shows that the noise is comprised mostly in the range of 150\,e and 200\,e, still compatible with pre-irradiation values.\\
Measurements with an $^{241}$Am $\gamma$-source have been performed to verify both the functioning of the bump-bond connections between the sensor and the front-end chip, and the front-end tuning. 
Figure \ref{fig:Am241} shows the $^{241}$Am $\gamma$-source spectrum obtained with a non-irradiated sensor biased at 350\,V.
Since the $\gamma$-source emits photons with an energy of 60\,keV that deposit a charge of  about 16.6\,ke in silicon, the position of the photoelectric peak of the measured charge distribution is used to verify the front-end calibration. 
The measured peak position at about 15\,ke agrees with the expectation within the uncertainty due to the ToT to charge calibration process. This uncertainty contains several components, and is estimated to be about 10$\%$. It is assigned to all charge measurements that follow.
\begin{figure}[h!]
\centering
\includegraphics[scale=0.45]{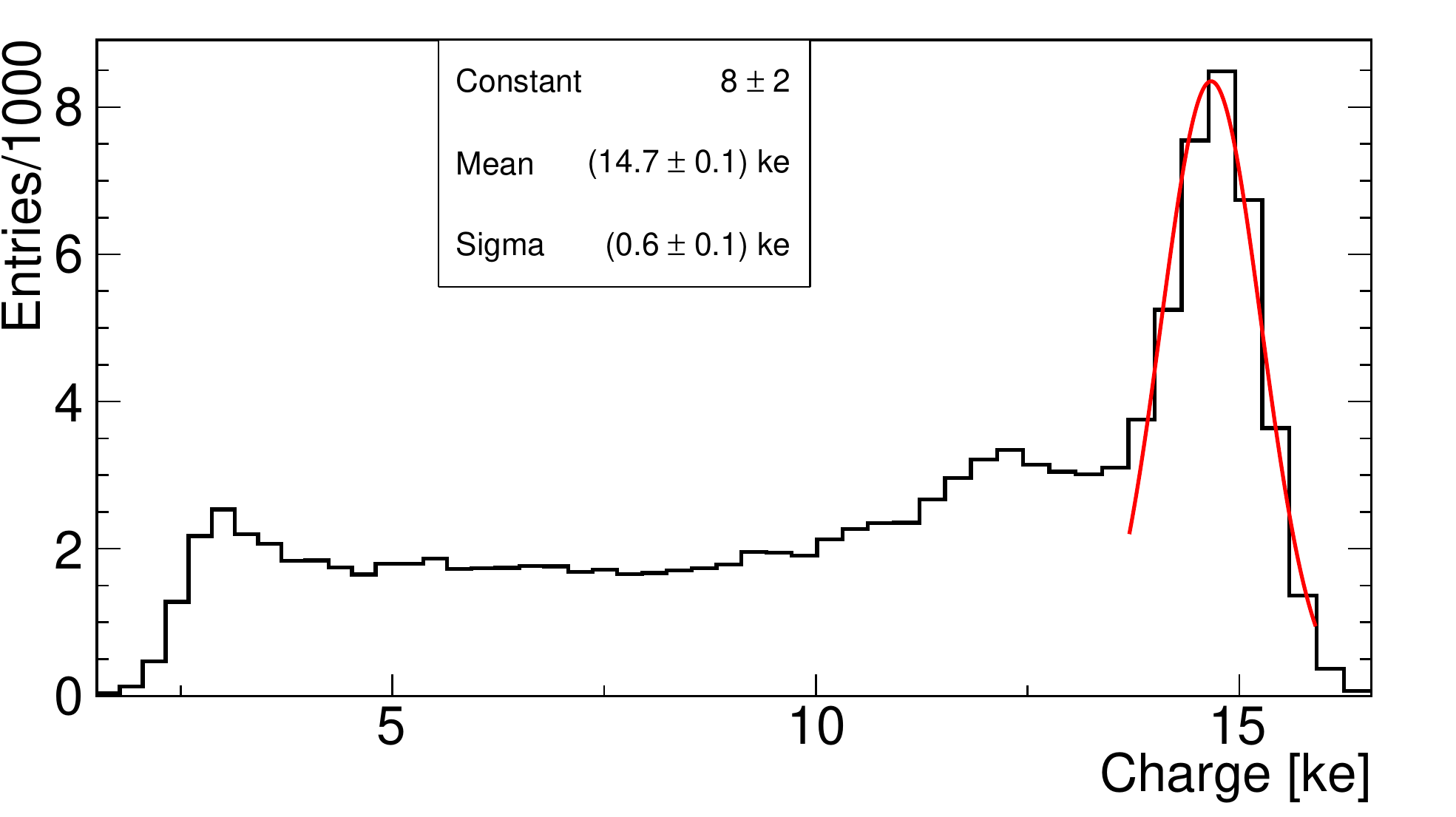}
\caption{$^{241}$Am spectrum measured with a non-irradiated module, biased at 350\,V, and kept at room temperature.}
\label{fig:Am241}
\end{figure}
\begin{figure}[h!]
\centering
\subfigure[]{
\includegraphics[scale=0.45]{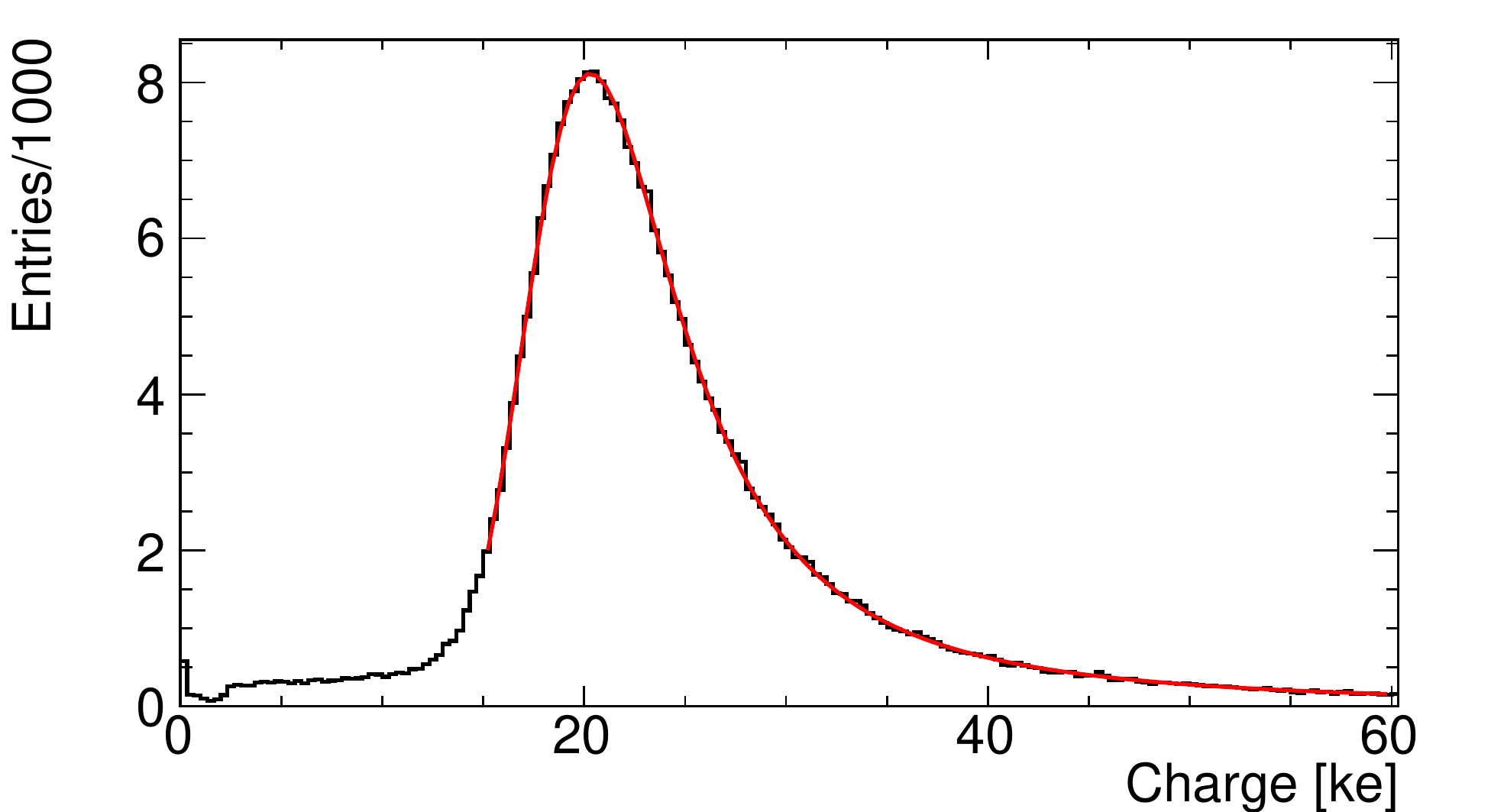}
\label{fig:unirrad-CCE}
}
\subfigure[]{
\includegraphics[scale=0.4]{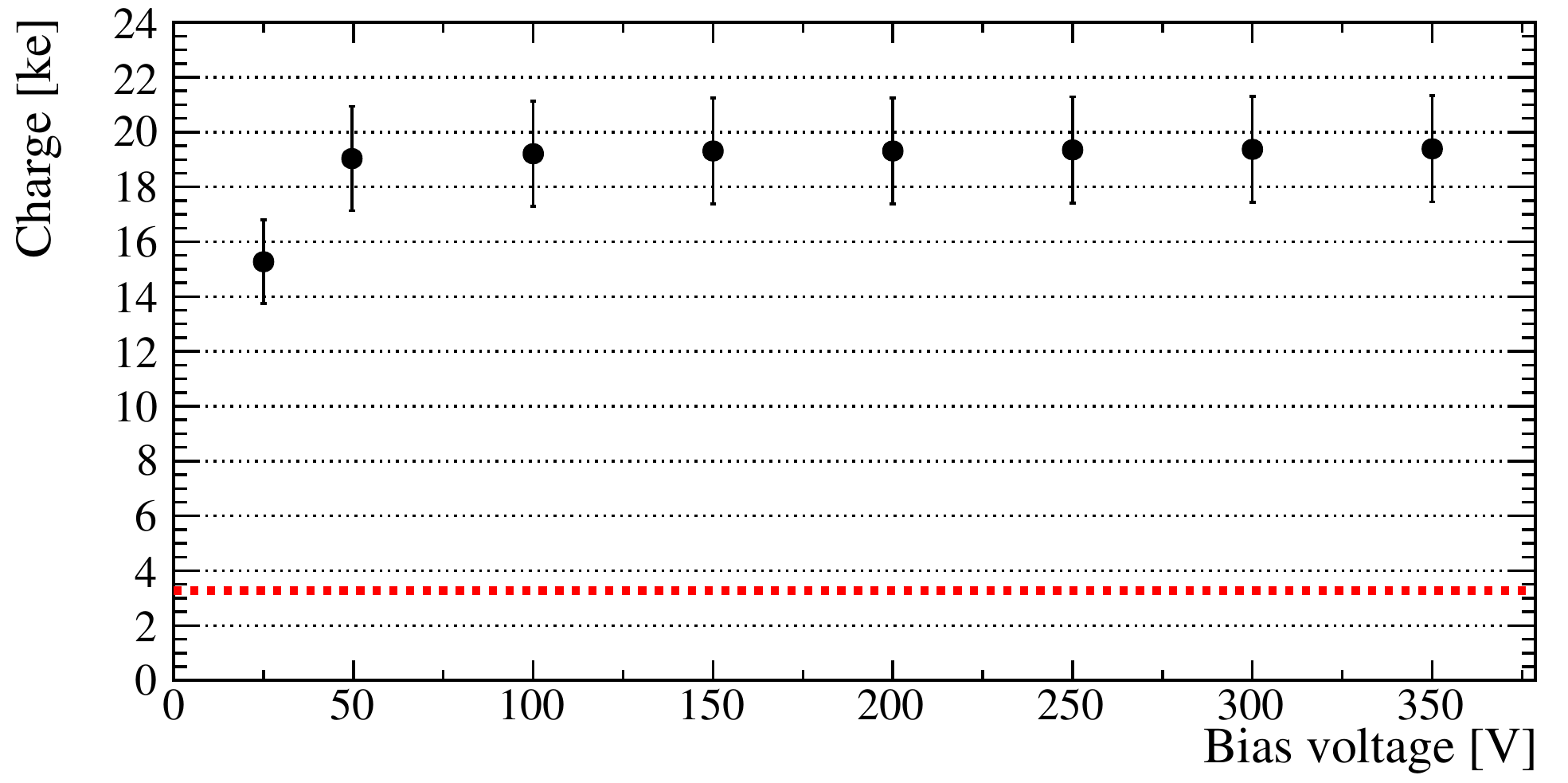}
\label{fig:unirrad-CCE_hv}
}
\caption{\subref{fig:unirrad-CCE} Charge collection distribution obtained from $^{90}$Sr  source measurement with a non-irradiated module, biased at a voltage of 150\,V, and kept at room temperature. The distribution refers to all cluster sizes and the measured most probable value (MPV) is (19$\pm$2)\,ke. \subref{fig:unirrad-CCE_hv} Dependence of the MPV on the applied bias voltage, obtained with $^{90}$Sr  source scans with a non-irradiated device. The dotted line denotes the value of the front-end threshold, tuned at 3200\,e. }
\end{figure}
\\
Using a $^{90}$Sr $\beta$-source, the charge collection has been mesured in a climate chamber at low humidity and stable air temperature ($+20\,^{\circ}\mathrm{C}$ for non-irradiated samples and between $-20\,^{\circ}\mathrm{C}$ and $-60\,^{\circ}\mathrm{C}$, with most of the measurements at $-50\,^{\circ}\mathrm{C}$, for irradiated samples). 
The reported temperatures are the measured air temperatures in the climate chamber. Using IV measurements these have been estimated to be 5\,$^{\circ}$C to 10\,$^{\circ}$C lower than the temperature on the sensor, where the dissipated power of the front-end chip leads to a temperature increase. 
The experimental setup used for the charge collection study consists of a $^{90}$Sr  source, placed on top of a scintillator attached to a photomultiplier. The modules were placed between the source and the scintillator. \\
Figure \ref{fig:unirrad-CCE} shows the distribution of the charge collection, obtained with a non-irradiated sample, biased at 150\,V. The most probable value (MPV), obtained from the signal of all cluster sizes, resulting from a convolution of a Landau and a Gaussian distribution, is (19$\pm$2)\,ke. 
The charge collection  was also studied as a function of the bias voltage. Figure \ref{fig:unirrad-CCE_hv} shows that the MPV is  constant for a bias voltages above 150\,V and decreases below the depletion voltage at about 60\,V.
\begin{figure}[h!]
\centering
\includegraphics[scale=0.45]{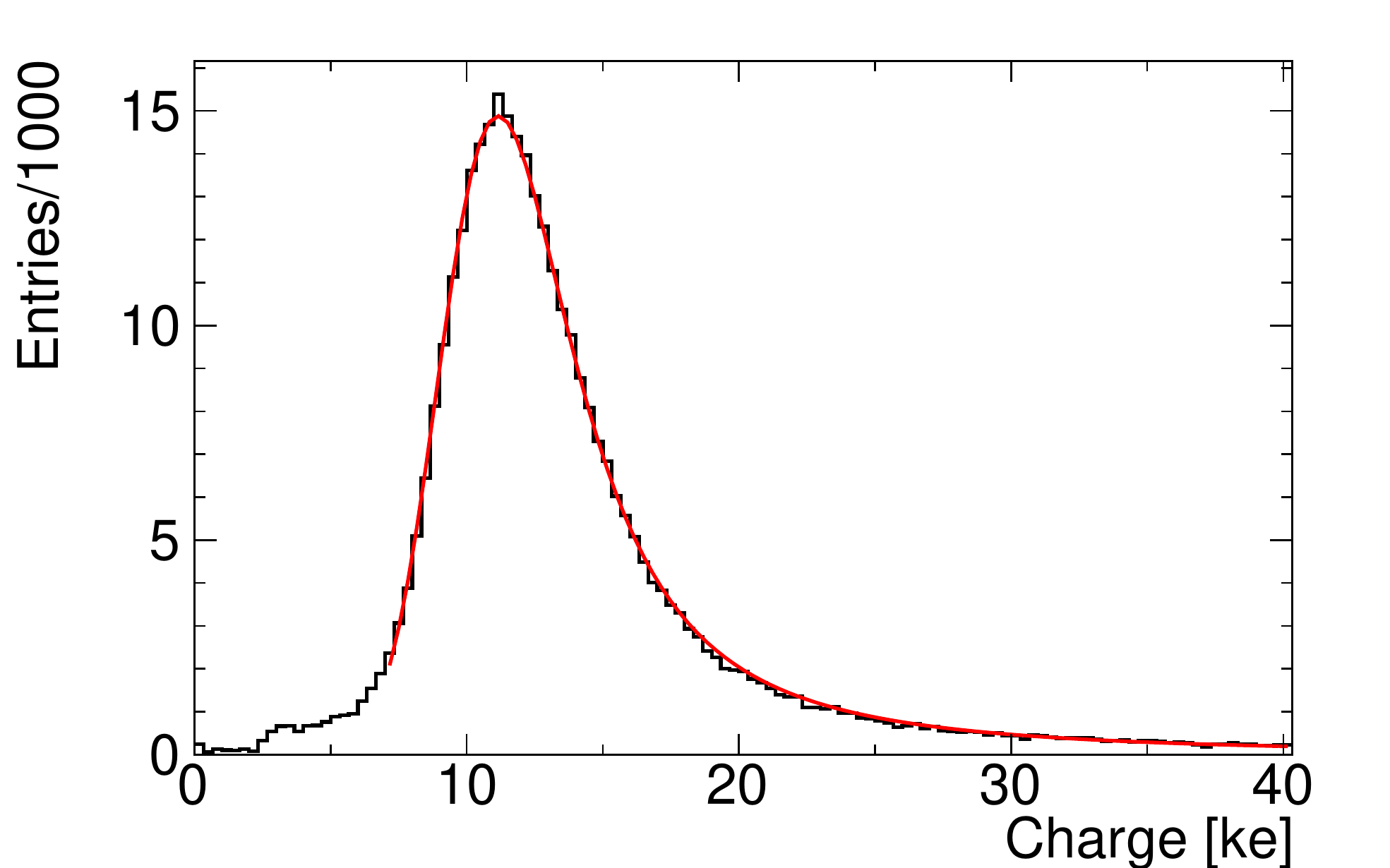}
\caption{Charge collection distribution obtained from a $^{90}$Sr  source measurement on sample E, irradiated up to 5\,x\,$10^{15}$\,n$_{\mathrm{eq}}$cm$^{-2}$, biased at 1000\,V, and kept at an environment temperature of  $-60\,^{\circ}\mathrm{C}$.The distribution refers to all cluster sizes, and the measured MPV is (11$\pm$1)\,ke.}
\label{fig:irrad-CCE}
\end{figure}
\\
The same characterization has been performed with irradiated samples. Figure \ref{fig:irrad-CCE} shows the distribution of the charge collection, obtained for sample E, irradiated at 5\,x\,10$^{15}$\,n$_{\mathrm{eq}}$cm$^{-2}$, biased at 1000\,V, and kept at an air temperature in the climate chamber of  about $-60\,^{\circ}\mathrm{C}$. Also in that case, the distribution refers to all cluster sizes, and the measured MPV is (11$\pm$1)\,ke. The degradation of the collected charge is attributed to the fact that the sensor is not fully depleted, and part of the charge released along the bulk is trapped, and thus not collected.\\
The summary of the charge collection measurements as a function of the bias voltage is reported in Figure \ref{fig:irrad-CCE-HV} for neutron irradiated modules, where the front-end chips were tuned to a threshold of 3.2\,ke (dotted line). As expected, for a given bias voltage, the MPV decreases with the received fluence \cite{RD50-1}.\\
The results obtained with the sample E have been compared with those from an n-in-n sample irradiated to the same equivalent fluence of 5x10$^{15}$\,n$_{\mathrm{eq}}$ \cite{N-in-N}. 
The two technologies show very similar behaviour of the charge collection efficiency.\\
\begin{figure}[h!]
\centering
\includegraphics[scale=0.44]{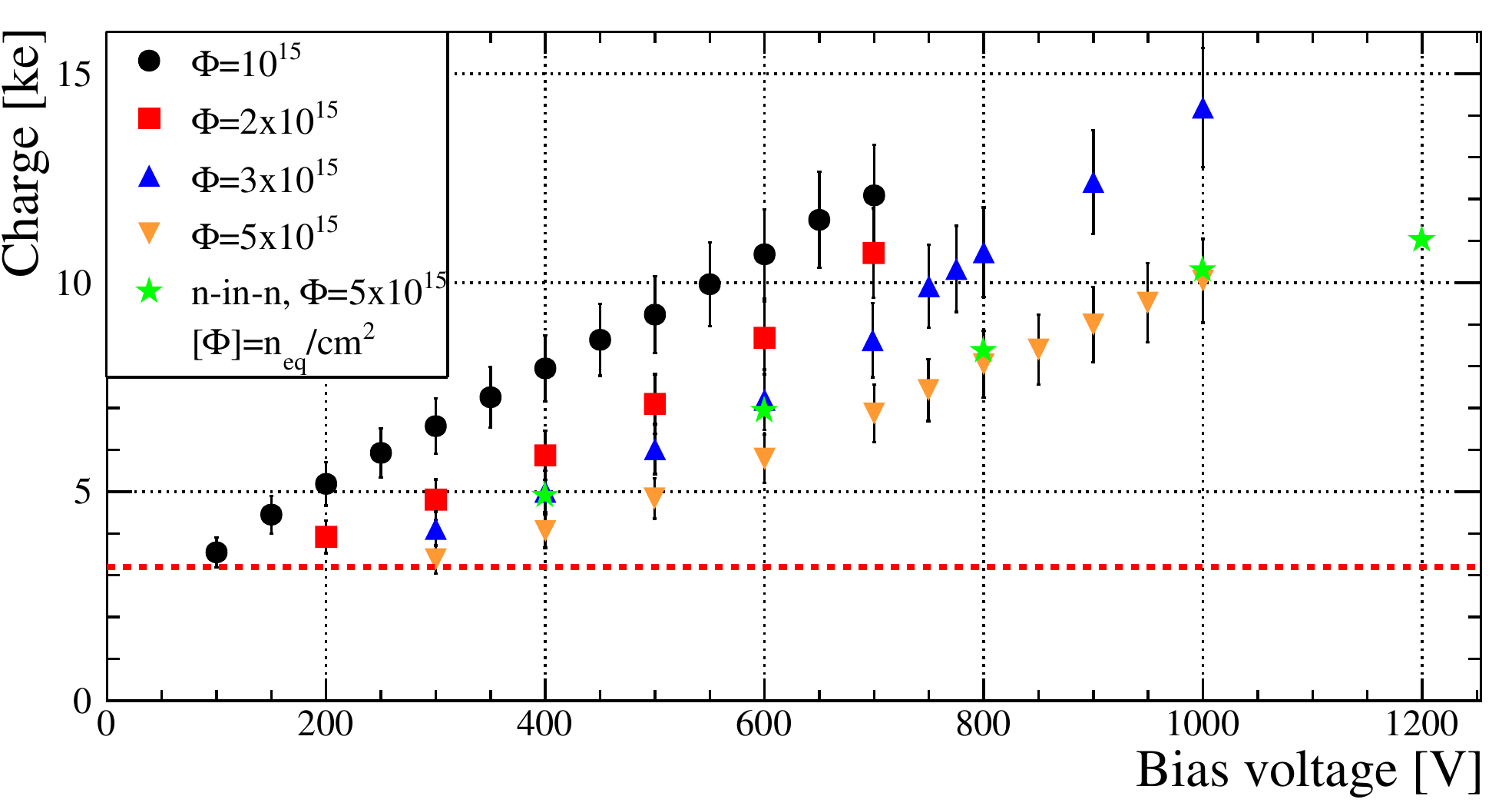}
\caption{Charge collected by neutron irradiated modules, obtained from $^{90}$Sr  measurements, as a function of the bias voltage. The modules were tuned to a front-end threshold of 3200\,e. The uncertainties shown are due to the spread of the internal calibration capacitors. The results from the n-in-n modules are from \cite{N-in-N}.}
\label{fig:irrad-CCE-HV}
\end{figure}

\section{Beam test measurements and results}
\label{sec:TB}
Beam test studies were conducted  at the CERN SPS/H6 beam-line with 120\,GeV/c $\pi^{+}$ using the EUDET telescope \cite{EUDET_2} for tracking. An overview of the test-beam setup can be found in \cite{WeingartenTB}. Cooling for the irradiated modules was achieved by using dry ice within an insulated volume. 
\subsection{Beam test analysis}
In the beam test analysis,  a fiducial region spanning 14 columns and 118 rows in the centre of the sensor is used. To compensate for movements due to temperature changes, the alignment procedure is repeated for each run. 	
\subsection{Tracking efficiency}
\begin{figure*}[th!]
\centering
\begin{minipage}[b]{\textwidth}
\centering
\subfigure[]{
\includegraphics[scale=0.9]{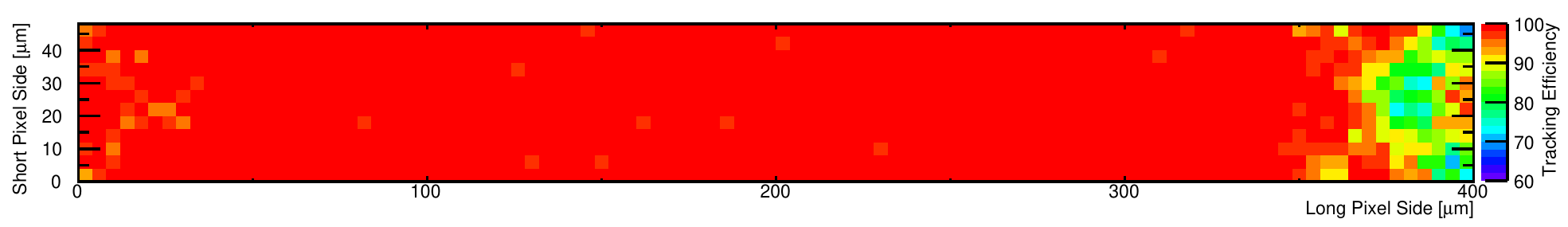}
\label{fig:tackeff}
}
\subfigure[]{
\includegraphics[scale=0.9]{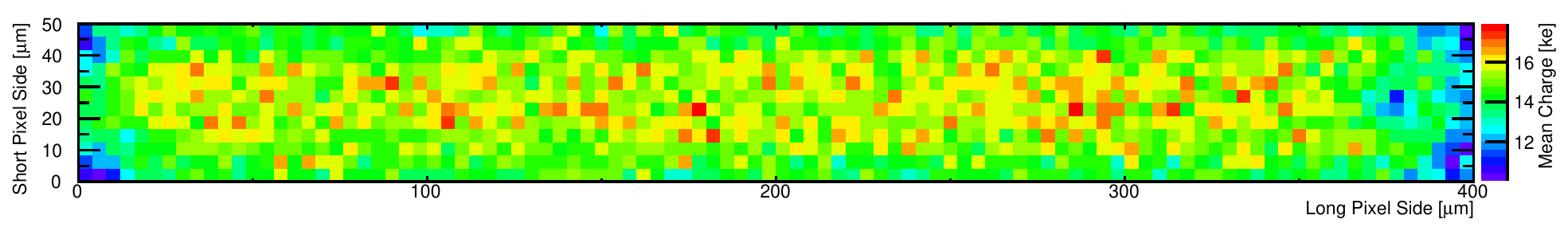}
\label{fig:chargemapsampleA}
}
\subfigure[]{
\includegraphics[scale=0.9]{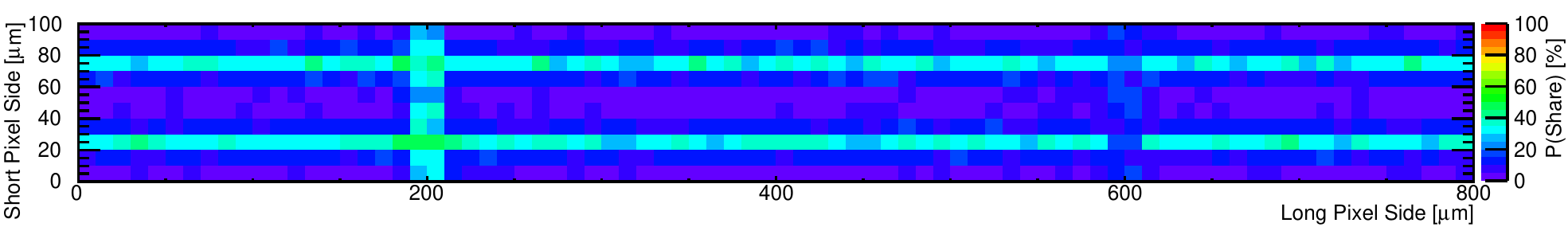}
\label{fig:chargesahring}
}
\subfigure[]{
\includegraphics[scale=0.96]{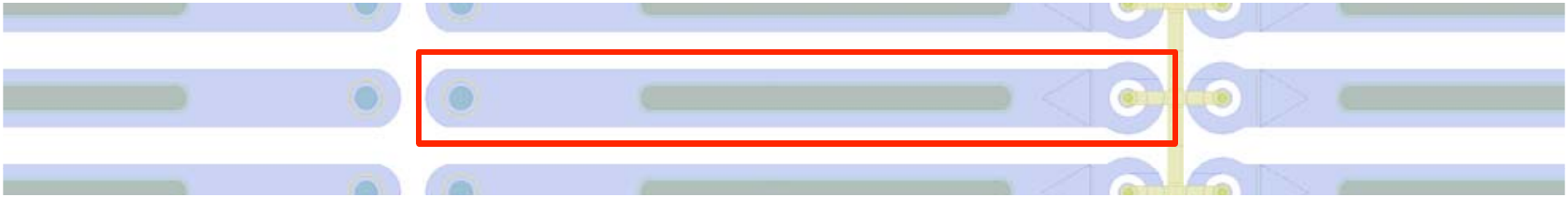}
\label{fig:Fig12d}
}
\caption{ 
Several quantities as functions of the impact point predicted by the telescope.
Shown are \subref{fig:tackeff} the mean tracking efficiency for module E, operated at a bias voltage of 600\,V, \subref{fig:chargemapsampleA} the mean collected charge for module A, operated at a bias voltage of 500\,V, and \subref{fig:chargesahring} the probability of charge sharing between pixels for module B, operated at a bias voltage of 700\,V.
The pixel geometry is shown in \subref{fig:Fig12d} both for an individual pixel as used in \subref{fig:tackeff}, \subref{fig:chargemapsampleA} indicated by the central frame, and for the two by two pixel matrix exploited in \subref{fig:chargesahring}.}
\end{minipage}
\end{figure*}

For the non-irradiated module as well as for module E, irradiated up to 5\,x\,10$^{15}$\,n$_{\mathrm{eq}}$cm$^{-2}$, tracking efficiencies were determined.
The tracking efficiency is calculated as the ratio of the number of clusters matched to impact points calculated from the telescope tracks, to all measured cluster in the respective module, within the fiducial region, omitting masked pixels and their corresponding neighbours. A track is matched to a cluster if their distance is less than 400\,$\mu$m in the long, and 150\,$\mu$m in the short pixel direction. 
For the non-irradiated module, a tracking efficiency of (99.3\,$\pm$\,0.2)\,\% is measured, with a homogenous behaviour over all the pixels. The efficiency for module E  irradiated up to 5\,x\,10$^{15}$\,n$_{\mathrm{eq}}$cm$^{-2}$ is still as high as (98.6\,$\pm$\,0.3)\,\%, when using a threshold of 3.2\,ke, and at bias voltage of 600\,V. Lowering the threshold to 2\,ke does not significantly change the efficiency. Figure\,\ref{fig:tackeff} shows the mean tracking efficiency, as a function of the impact point predicted by the telescope, within a pixel geometry indicated by the central frame in Figure\,\ref{fig:Fig12d}. While in the main pixel region the efficiency is 99.8\,\%, in the region of the punch through biasing, the efficiency is lower, due to the implant structure.
%
%

\subsection{Charge collection}
The position of the charge released within the pixel can be determined from the reconstructed impact point, which has a resolution of 2.5\,$\mu$m \cite{WeingartenTB}, as given by the precision of the telescope. 
Figure\,\ref{fig:chargemapsampleA} shows the mean collected charge, as a function of the impact point predicted by the telescope, within a pixel geometry as indicated by the central frame in Figure\,\ref{fig:Fig12d}, for module A, irradiated up to 1\,x\,10$^{15}$\,n$_{\mathrm{eq}}$cm$^{-2}$, and biased at 500\,V. As in the non-irradiated modules, the bias dot is visible. 
Additionally, for irradiated modules, the edge and corner regions have lower charge collection efficiencies, since charges are shared more likely between more pixels. Clearly, for the entire pixel area, the MPV exceeds twice the threshold of 3200\,e.
%
\subsection{Charge sharing}
\begin{figure}[h!]
\centering
\includegraphics[scale=0.4]{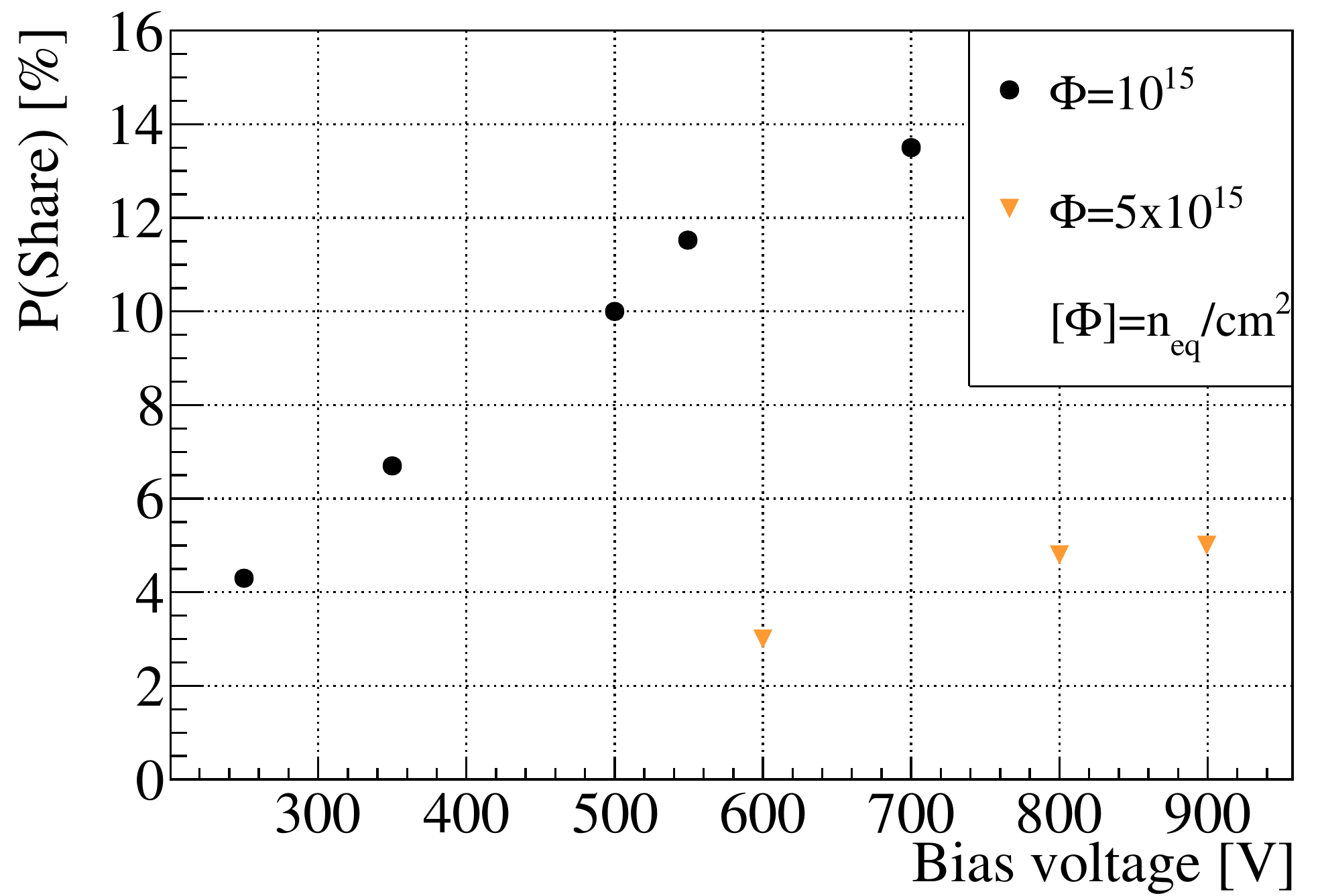}
\caption{Dependence of charge sharing probability as  function of the bias voltage for module B, irradiated up to 1\,x\,10$^{15}$\,n$_{\mathrm{eq}}$cm$^{-2}$, and E, irradiated up to 5\,x\,10$^{15}$\,n$_{\mathrm{eq}}$cm$^{-2}$, within the testbeam.}
\label{fig:chargesahringvsbias}
\end{figure}

The position resolution of the detector is improved by taking charge sharing between different pixels into account. On the other hand, if the charge measured in one pixel is below threshold, due to the charge sharing, the tracking efficiency will be diminished. 
In the beam test, the probability for a charge to be shared with other pixels can be determined as a function of the track impact point. 
In Figure \ref{fig:chargesahring}, the charge sharing probability, P(share), is shown for module B, irradiated up to 1\,x\,10$^{15}$\,n$_{\mathrm{eq}}$cm$^{-2}$, operated at a bias voltage of 700\,V. It is defined as the fraction of hits sharing charge with their corresponding neighbours,  to all hits at this position. 
In the figure, showing a two by two pixel matrix centered on one pixel, all data from the module is superimposed.
The orientation of the central pixel is the same as for figure\,\ref{fig:pixel-layer}, with the bias dot on the right hand side. As expected, the sharing probability is highest at the side opposite to the bias dot, and in between four pixels, i.\,e. at $(x,y)\in\{(200\,\mu$m$, 25\,\mu$m), ($200\,\mu$m, $75\,\mu$m)\}. On the punch through biasing side, i.\,e. at $x=600\,\mu$m, the charge sharing is reduced, since, when distributed over more than one pixel, in this area the collected charge more likely does not exceed the threshold.
The other zones of high sharing probability align naturally with the borders of the pixel cell. 
The dependence of the overall charge sharing probability on the applied bias voltage is shown in Figure \ref{fig:chargesahringvsbias}. Due to the higher charge collection efficiencies at higher voltages, and thus increased likelihood of the collected charge to exceed the threshold also in the neighbouring pixel, the probability rises with bias voltage. The collected charge decreases with the fluence, and the probability that a hit is below threshold, and not detected, increases accordingly, leading to a lower sharing probability.
If the threshold is lowered from 3.2\,ke to 2\,ke for module E, irradiated at 5\,x\,10$^{15}$\,n$_{\mathrm{eq}}$cm$^{-2}$, and biased at 600\,V, the overall charge sharing probability increases to 5\,\%.

\section{Conclusion}
\label{sec:conclusion}

In view of the LHC upgrade phases towards HL-LHC the ATLAS experiment plans to upgrade the Inner Detector  with an all silicon system. The n-in-p silicon technology is a promising candidate for the pixel upgrade thanks to its radiation hardness and cost effectiveness, that allow for enlarging the area instrumented with pixel detectors.\\
In this paper, the properties of novel n-in-p planar pixel detectors were presented, and a signal over threshold of more than three, together with low noise, has been obtained with a module irradiated up to 5\,x\,$10^{15}\,$n$_{\mathrm{eq}}$cm$^{-2}$. At the maximum fluence, a charge of about 10\,ke was collected, at a bias voltage of 1000\,V, with the discriminator threshold tuned to 3200\,e. 
Thanks to the coating of the sensor surface with a BenzoCycloButene layer, the modules can be stably operated at these high voltages, i.e.~over days of operating time, no sparks between the front-end chips and the sensors were observed.
For the module irradiated at the maximum fluence of 5\,x\,10$^{15}$\,n$_{\mathrm{eq}}$cm$^{-2}$, the tracking efficiency is  (98.6\,$\pm$\,0.3)\,\%, when setting a threshold of 3200\,e, and biasing the module at 600\,V.\\
A next n-in-p pixel sensor production is on-going at CiS, on 4" wafer of 150\,$\mu$m, 200\,$\mu$m, and 300\,$\mu$m thickness. These pixel sensors have a geometry compatible with the new ATLAS FE-I4 chip \cite{FEI4}, with a pixel dimension of 50\,x\,250\,$\mu$m$^{2}$. The width of the inactive region is reduced to 450\,$\mu$m per side. The reduced bulk thickness should lead to an enhanced charge collection efficiency after irradiation, as observed in many studies performed within the RD50 collaboration.

\section{Acknowledgement}
This work has been partially performed in the framework of the CERN RD50 Collaboration.
The authors would like to thank A. Dierlamm (KIT), for performing the 25\,MeV proton irradiation in Karlsruhe (Germany),  and V. Cindro (JSI), for performing the neutron irradiation in Ljubljana (Slovenia). Part of these irradiations were performed within the framework of the Helmholtz alliance.\\
A big thank to the ATLAS PPS testbeam crew\footnote{The ATLAS PPS testbeam group members: S.~Altenheiner, M.~Backhaus, M.~Beimforde, M.~Benoit, M.~Bomben~(coordinator -2011), G.~Calderini, D.~Forshaw, Ch.~Gallrapp, M.~George, S.~Gibson, S.~Grinstein, J.~Idarraga, Z.~Janoska, J.~Jansen, J.~Jentsch, O.~Jinnouchi, T.~Kishida,  A.~La~Rosa, S.~Libov, T.~Lapsien, A.~Macchiolo, G.~Marchiori, D.~M\"unstermann, R.~Nagai, C.~Nellist, G.~Piacquadio, B.~Ristic, I.~Rubinsky, A.~Rummler, Y.~Takubo, G.~Troska, S.~Tsiskaridze, I.~Tsurin, Y.~Unno, P.~Weigell, J.~Weingarten~(coordinator -2010) and T.~Wittig.}, the EUDET people, and the CERN SPS North Area teams for their support. Additional thanks to I. McGill (CERN) for the wire-bonding of the irradiated modules.

\end{document}